# KUIPER BINARY OBJECT FORMATION

R. C. Nazzario, K Orr, C. Covington, D. Kagan, and T. W. Hyde
Center for Astrophysics, Space Physics and Engineering Research,  Baylor University, Waco, TX 76798-7310, USA, Truell_Hyde@Baylor.edu/ phone: 254-710-3763

## ABSTACT

It has been observed that binary Kuiper Belt Objects (KBO's) exist contrary to theoretical expectations. Their creation presents problems to most current models. However, the inclusion of a third body (for example, one of the outer planets) may provide the conditions necessary for the formation of these objects. The presence of a third massive body not only helps to clear the primordial Kuiper Belt but can also result in long lived binary Kuiper belt objects. The gravitational interaction between the KBO's and the third body causes one of four effects; scattering into the Oort cloud, collisions with the growing protoplanets, formation of binary pairs, or creation of a single Kuiper belt object. Additionally, the initial location of the progenitors of the Kuiper belt objects also has a significant effect on binary formation.

## 1. INTRODUCTION

Observations of the Kuiper Belt indicate that a larger than expected percentage of KBO's (approximately 8 out of 500) are in binary pairs. The formation of such objects presents a conundrum (Funato et al., 2004). Two competing theories have been postulated to try to solve this problem. One entails the physical collision of bodies (Weidenschilling 2002) while the other utilizes dynamical friction or a third body to disappate excess momentum and energy from the system (Goldreich et al. 2002). In general, binaries also tend to differ significantly in their masses, such as the Earth-Moon system and asteroid binary systems (Trujillo, 2003). However, the binaries in the Kuiper Belt tend to be of similar size (Bernstein et al., 2004; Altenhoff et al., 2004). This paper numerically investigates a gravitational formation mechanism employing a Neptune size protoplanet to explain the possible formation of similar sized KBO binaries and their subsequent evolution. Section 2 details the numerical model while Section 3 discusses the initial conditions. Finally, in Section 4 the results are discussed with Section 5 containing the conclusions.

## 2. NUMERICAL MODEL

The numerical method employed in this work is a $5^{th}$ order Runga-Kutta algorithm (Nazzario, 2002) based on the Butcher's scheme (Chapra and Canale, 1985) with a fixed time step. In this work, time step of 2 days was used in order to reduce truncation and round-off errors yet still yield reasonable CPU run times with a simulation time of 1000 years observed. The forces considered in this model include the gravitational attraction of the Sun, the 7 major planets (Venus through Neptune) and all other KBO's. The corresponding accelerations due to these forces are given by Eq. (1).

$$\vec{a}_i = -\frac{GM_{Sun}}{r^2}\hat{r} - \sum_{i=1}^{n}\sum_{\substack{j=1\\i\neq j}}^{n}\frac{Gm_j}{r_{ij}^2}\hat{r}_{ij} \tag{1}$$

Particles are considered removed from the system when they venture beyond 100 AU or enter the sphere of influence (SOI) of one of the 7 major planets. The SOI is defined (as shown in Eq. (2)) as the distance from the secondary (in this case the planet) where bodies interact with one another as opposed to be being primarily acted upon by the Sun. The SOI is given by

$$r_{SOI} \approx \left(\frac{M_{Secondary}}{M_{Primary}}\right)^{2/5} r \tag{2}$$

where $M_{Secondary}$ is the mass of planet, $M_{Primary}$ is the mass of the Sun, and r is the distance of separation between the primary and secondary bodies. Collisions between objects are also allowed and consequently their size is allowed to change.

3. INITIAL CONDITIONS

The Kuiper Belt is divided into three distinct populations; the classic Kuiper Belt Objects, the resonant KBO's, and the scattered KBO's. The simulations presented here are representative of the classic Kuiper Belt population (containing 50% of all KBO's) where the eccentricities are small (less than 0.2) and in the ecliptic plane (Trujillo 2003). This KBO population has a size distribution for which the number of 100 km particles is greatest (Bernstein et al., 2004), therefore; a radius of 100 km was chosen for the radius of simulated KBO's. A mean density of 3.5 g/cm$^3$ was given to these objects (corresponding to dirty water ice) yielding individual KBO masses of 1.4x10$^{19}$ kg.

Two primary simulations were conducted. The first involved starting with a complete ring of KBO's (Fig. 1) while the second initially employed a higher surface number density ring arc of KBO's (Fig. 2). A total of 4500 KBO's were initially employed in the ring simulations with particles evenly distributed between 20 and 33 AU. In the ring arc simulations, 2700 KBO's were included with particles distributed between 27 and 36 AU in a 45° arc starting approximately 60° ahead of the orbit of Neptune (Fig. 2). Particles were initially given the velocities they would have had if they were in circular orbits.

4. RESULTS

Binaries were found to form only rarely in both simulations. Simulations where particles were initially placed in a ring structure (using 4500 test particles) resulted in the total formation of 14 binaries. The arc simulations, while using far fewer particles (only 2700 test particles), resulted in a total creation of 276 binary pairs. Virtually all of these binaries were transients lasting less than 50 years with only 5 instances where binaries pairs were shown to be stable for at least 100 years. Transient binaries first started forming after 400 years of simulation time with stable binaries appearing 750 years into the simulation.

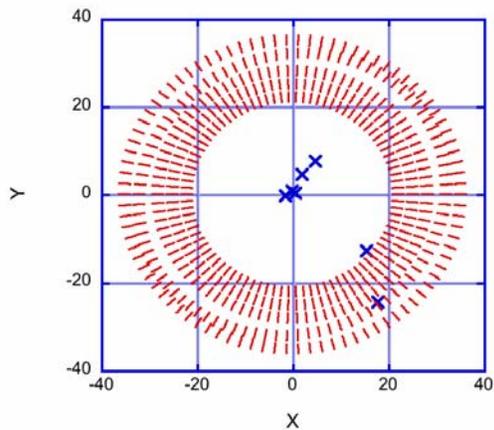 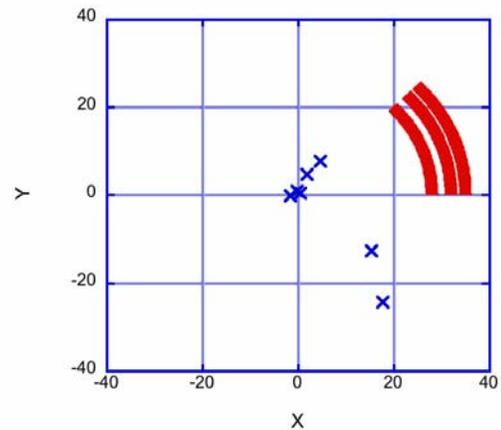

Figure 1. Initial positions of particles started in a ring. 'X's represent the planets and dots represent the KBO Objects

Figure 2. Initial positions of particles started in a ring arc. 'X's represent the planets and dots represent the KBO Objects

Arc simulations showed multiple areas of enhanced concentration, which also encouraged binary formation as well as the appearance of retrograde orbits, which are inherently more unstable than are prograde orbits which can also result. Figures 5 and 6 show two such distinct populations for each set of simulations.

Finally, multiple particle systems consisting of up to 4 objects in orbit about a common center were also observed in arc simulation data. These were again shown to be quasi-stable as they persisted less than 50 years before breaking apart or exchanging partners with other objects.

Collisions between objects were allowed to occur. However, no collisions occurred between KBO's over the lifetime of the simulation. Neptune (the closest planet to the KBO's) was responsible for removing 267 KBO's (6% of the total number of KBO's) during the ring simulation. In the arc simulation 305 KBO's were removed by Neptune, corresponding to 11% of the number of KBO's simulated. Also, during the simulation no KBO's were ejected into the Oort cloud.

5. CONCLUSIONS

Presently, there are only 8 known KBO binaries in the Kuiper Belt (approximately 4%). These simulations indicate binary formation rates of only 0.3% for simulations with particles initially placed in a full ring but show that increasing the surface number density of such particles can result in an approximately 10% formation rate from particles placed initially in ring arcs. These percentages are in good agreement with Goldreich et al. (2002) who predicted 5% of KBO's should be binaries (using gravitational interactions). Although many of these KBO binaries show only short-term stability, this can be correlated to previous results for the lifetimes of binaries (Petit and

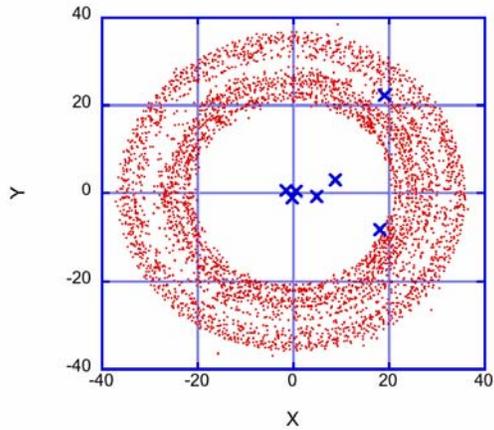

Figure 3. Final positions for particles started in a ring arc. 'X's represent the major planets while dots represent the KBO objects.

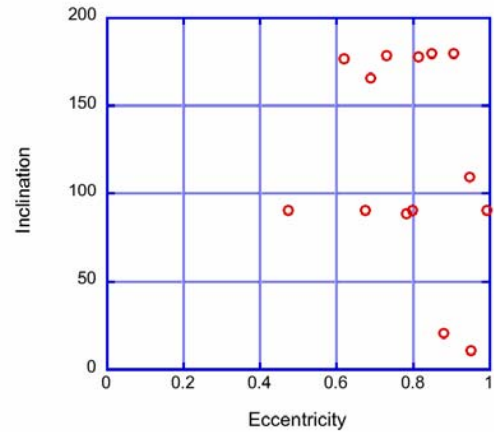

Figure 5. Eccentricity-inclination plot for KBO objects which became binaries after being started in a complete ring. Open circles represent individual KBO's.

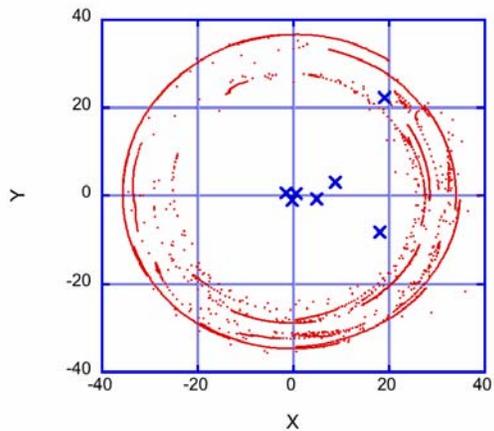

Figure 4. Final positions for particles started in a ring arc. Blue dots represent the major planets while dots represent the KBO objects.

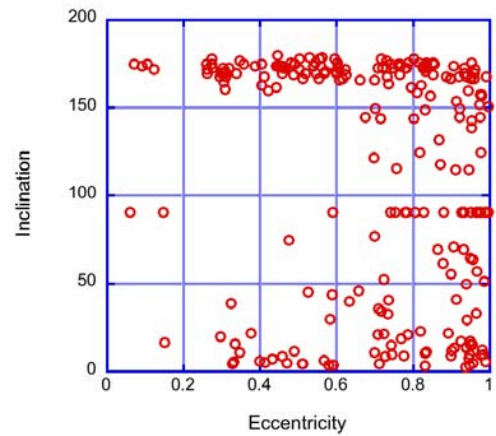

Figure 6. Eccentricity-inclination plot for KBO objects which became binaries after being started in a ring arc. Open circles represent individual KBO's.

Mousis, 2004) who found that over the age of the solar system at least 1/3 of all binaries would be disrupted. This is also in agreement with Nesvorny and Dones (2002) who found that only 50% of KBO binaries survived.

The finding of systems consisting of greater than 2 objects also agrees with Goldreich et al. (2002) with they postulated that systems of higher multiplicity could occur.

The concentration of KBO particles in a given region must be relatively high near the position of a large planet for even transient binary KBO formation. KBO densities on the order of $12/AU^2$ resulted in transient binaries being formed while binary formation was found to be negligible for KBO densities lower than approximately $2/AU^2$. This is in agreement with predictions that the Kuiper Belt was originally up to 100 times as massive as it is presently. The higher density will also enable Neptune to accrete many more objects contributing to its orbital migration.